\documentstyle[12pt]{article}
\begin{document}
\title{
\begin{flushright}
{\small Talk presented at XIII International Seminar 
"Relativistic nuclear physics and Quantum Chromodynamics", 
 Dubna, September 1996} \end{flushright} \vspace{0.5cm} Bosonized 
 Formulation of Lattice QCD.  \author{ A.A.Slavnov \thanks{e-mail: 
 slavnov@class.mian.su} \\ Steklov Mathematical Institute, Russian Academy of 
 Sciences,\\ Vavilov st.42, GSP-1,117966, Moscow, Russia }} \maketitle

\begin{abstract}

 Problems in lattice gauge models with fermions are discussed.
A new bosonic Hermitean effective action for lattice QCD with dynamical 
quarks is presented.  In distinction of the previous versions \cite{AS1}, 
\cite{AS2} it does not include constraints and is better suited for 
Monte-Carlo simulations.  \end{abstract} \section {Problems in lattice 
QCD}

Lattice formulation of QCD seems at present to be the only regular method 
to study nonperturbative effects in this model. Wilson's formulation of 
lattice Yang-Mills theory \cite{W} provides a solid basis for computations 
in pure gluodynamics. However real calculations pose certain requirements 
on the size and spacing of a lattice. To supress finite volume effects the 
size of a lattice should be bigger than typical hadron diameter, which 
gives the estimate $L \geq 3 fermi$. The experience shows that on such 
lattices many physical effects can be described with sufficient accuracy 
for lattice spacing $ \leq 0.1 fermi$.

Modern computer facilities allow to concider the lattices of the size 
$32^4$ (record size is $64^4$), which seems to be sufficient for reliable 
calculations in QCD without quarks or in the quenched sector neglecting 
dynamical quark loops. In this way reasonably good results were 
obtained in calculation of hadron spectrum and form- \\factors, 
studying QCD at finite temperature and density.

However, contrary to the 
comon lore that "all difficulties in gauge theories are related to 
Yang-Mills sector and inclusion of matter fields makes no problem", in 
lattice QCD the situation is different.  Whereas the pure gauge sector is 
well understood, treating of dynamical fermions is far from satisfactory 
and both principal and technical problems are still present.

Well known problem of fermion spectrum degeneracy on the lattice makes 
difficult to consider chiral fermions as well as to study chiral effects 
in QCD (For detailed discussion and references see e.g. reviews \cite{MG}, 
\cite{DP}). Recently a new formulation of chiral fermion models on the 
lattice was proposed \cite{ASch1}, \cite{ASch2}, which allows to remove 
unwanted fermon doublers preserving chiral invariance of the theory at 
least in the framework of weak coupling expansion. Another approach, 
pretending on nonpertur-\\bative solution of the problem was proposed in 
papers \cite{NN1}. 

However to estimate a real value of all 
these proposals extensive nonperturbative calculations are needed. Here 
comes a major technical problem, which is common for all models with 
dynamical fermions, in particular QCD. Calculation of physical quantities 
on the lattice can be done by computing path integrals which are 
approximated by corresponding finite sums. But in fermion models one 
encounters path integrals over anticommuting variables which cannot be 
treated in this manner. 

The action we are interested in is usually
quadratic in Fermi fields and the corresponding integral is 
Gaussian. It is equal to the determinant of the quadratic form in the 
action, whereas integration over Bose fields produces the inverse 
determinant. Hence the most direct way to compute a fermion 
detreminant is to calculate firstly the corresponding bosonic determinant
 and then to invert it. Alternatively one can perform integration over 
{}Fermi fields analytically to get a nonlocal bosonic effective action, thus 
avoiding integration over anticommuting variables. This procedure includes 
inversion of huge matrices and is very costly and time consuming. However 
considerable progress in this direction has been done, mainly using hybrid 
Monte-Carlo algorithm \cite{DKPR}. (For recent review of this and other 
methods see \cite{KJ}).

An alternative algorithm was formulated by M.L\"uscher \cite{ML}, 
\cite{ML1}, who proposed to calculate a fermion determinant replacing it 
by an infinite series of boson deter-\\minants. This method is based on 
the equation of the type \begin{equation} \det(Q)^2= \lim_{n \rightarrow 
\infty} [ \det(P_n(Q^2))]^{-1} \label{1} \end{equation} where $P_n(s)$ is 
a sequence of polynomials such that \begin{equation} \lim_{n \rightarrow 
\infty}P_n(s)=1/s, \quad 0<s \leq 1 \label{2} \end{equation} By cutting 
the sequence at some finite $n=N$ one gets an approximate value of fermion 
determinant expressed in terms of integrals over $N$ bosonic fields. This 
idea was developped further in papers \cite{BF}, \cite{BFG}.  It appears 
however that computational efforts needed in this approach are comparable 
with hybrid Monte Carlo algorithm mentioned above.
 
Another possibility was discussed recently in our papers \cite{AS1}, 
\cite{AS2}. In our approach a $D$-dimensional lattice fermion determinant 
was presented as a path integral of exponent of $D+1$-dimensional 
constrained bosonic effective action. This represen-\\tation is exact in 
the limit when the size of extra dimension becomes infinite and lattice 
spacing zero. For a finite lattice correction terms are present, whose 
value depends on the particular choice of effective bosonic action. In 
this talk I shall use the freedom in the choice of effective action to 
formulate a new version of the bosonization algorithm for lattice 
QCD  which seems to be better suited for Monte Carlo simulations, 
in particular it allows to get rid off the constraint equation.

\section{Bosonic effective action for lattice QCD}

We consider Euclidean QCD on the four-dimensional qubic lattice with 
spacing $a$.  The quark fields are four-dimensional spinors $ \psi^i(x)$, 
$i$ being a flavour index, $U_{ \mu}(x)$ is a gauge field. Let us 
introduce bosonic fields $ \phi(x,t)$ defined on a five-dimensional 
lattice, which have the same spinorial and internal structure as $ \psi$. 
The fifth component $t$ is defined on the one-dimensional chain of the 
length $L$ with the lattice spacing $b$
\begin{equation}
L=Nb, \quad 0 \leq n<N
\label{3}
\end{equation}
The effective bosonic action may be written as follows
 \begin{equation}
S=S_W(U)+a^4 \sum_x \{b \sum_{n=0}^{N-1} \{ [b^{-2}( \phi^*_{n+1}(x) 
\phi_n(x)+ \phi^*_n(x) \phi_{n+1}(x)-2 \phi^*_n \phi_n) - \label{4} 
\end{equation}
 $$
 -i[\phi^{*}_{n+1}(x) \gamma_5 \hat{D} \phi_n(x)- h.c.][b]^{-1}+ $$ 
$$ + \phi^{*}_n(x)D^2 \phi_n(x) -{ \sqrt{L}}( \phi^*_n(x)(m-
i \hat{D} \gamma_{5}) \chi(x)e^{-mbn}+h.c. \}+$$ $$+ \frac{L}{2m} 
\chi^*(x) \chi(x) \}.  $$ Here $S_W(U)$ is the standard Wilson action for 
lattice Yang-Mills field \cite{W}, $ \chi(x)$ are four-dimensional lattice 
bose fields having the same spinorial and internal structure as $ \psi$.  
 Lattice covariant derivative is denoted by 
$D_{ \mu}$ \begin{equation} D_{\mu} \psi(x)= \frac{1}{a}[U_{\mu}(x) 
\psi(x+a_{\mu})- \psi(x)] \label{4a} \end{equation} \begin{equation} 
\hat{D}=1/2 \gamma_{ \mu}(D_{ \mu}^*-D_{ \mu}) \label{5} \end{equation} 
(For simplicity we consider naive fermions, but all the construction is 
extended in a straightforward way to Wilson fermions).

The effective action (\ref{4}) allows to calculate in a standard way any 
gauge invariant correlation function. In particular we shall show that the 
square of the quark determinant can be presented as the following path 
integral
 \begin{equation}
\int \exp \{a^4 \sum_{i=1}^2 \sum_x \bar{\psi}_i(x)(
\hat{D}+m) \psi_i(x) \}d \bar{ \psi}d \psi = \lim_{L 
\rightarrow \infty, b \rightarrow 0} \int \exp \{-S \}d 
\phi^*_n d \phi_n d \chi^* d \chi \label{6} 
\end{equation}
where free boundary conditions in $t$ are assumed
\begin{equation}
\phi_n=0, \quad n<0, \quad n \geq N
\label{7}
\end{equation}
{}For finite $b$ and $L$ this equality has to be corrected by the terms of 
order $ O(b/a)^2, \\ O( \exp \{-mL \})$.  We consider the case of two 
flavours to provide positivity of the effective bosonic action. An 
analogous representation can be written for the modulus of quark 
determinant.

To prove the equation ( \ref{6}) we consider the following integral 
  \begin{equation} I= \int \exp \{a^4 
\sum_x \{b \sum_{n=0}^{N-1}  [b^{-2}( \phi^*_{n+1}(x) \exp \{-i 
\gamma_5 \hat{D}b \} \phi_n(x)+ h.c.-2 \phi^*_n \phi_n)  -
\label{8}
\end{equation}
  $$ - L^{1/2}( \phi^*_n(x)(m-i \hat{D} \gamma_5) \chi(x)+h.c. 
) \exp \{-mbn \}] $$
$$
+ \frac{L}{2m} \chi^*(x) \chi(x)\} \} d \phi^*_nd \phi_nd \chi^*d 
\chi .  $$ 
{}For small $b$ the exponent in the r.h.s. of eq.( \ref{8}) coincides with 
the effective quark action ( \ref{4}). Indeed taking into account that the 
operator $ \gamma_5 \hat{D}$ is bounded, $|| \gamma_5 \hat{D}|| \leq 
8a^{-1}$ we can approximate $ \exp \{-i \gamma_5 \hat{D}b \}$ by it's 
Taylor series. Keeping the first three terms we get exactly the action 
( \ref{4}). The remaining terms produce corrections of order $O(b^2/a^2)$, 
which can be done small by choosing appropriately lattice spacing in 
the extra dimension.

The operator $ \gamma_5 \hat{D}$ is Hermitean and it's 
eigenvalues are real.  R.h.s.  of eq.  (\ref{8}) can be written in 
the basis formed by the eigenvectors of the operator $ \gamma_5 
\hat{D}$:  \begin{equation} I= \int 
\exp \{ \sum_{\alpha} \{b \sum_{n=0}^{N-1} [(b^{-2})( 
\phi^{\alpha*}_{n+1} \exp \{-i D^{ \alpha}b \}\phi_n^{\alpha}+ h.c.
 -2 \phi^{\alpha*}_n \phi^{\alpha}_n)-
\label{9}
\end{equation}
 $$ - L^{1/2}( \phi^{\alpha*}_n (m-iD^{\alpha}) \chi^{\alpha}+ h.c.) \exp 
\{-mbn \}] $$
$$
+ \frac{L}{2m} \chi^*_{\alpha} \chi_{\alpha} \} \}d \phi^{\alpha*}_nd 
\phi^{\alpha}_nd \chi^{\alpha*}d \chi^{\alpha} $$ 

To calculate the integral (\ref{9}) we make the following change of 
variables:  \begin{equation} \phi_n^{\alpha} \rightarrow \exp 
\{-iD^{\alpha}nb \}\phi_n^{\alpha}, \quad \phi_n^{\alpha*} 
\rightarrow \exp \{iD^{\alpha}nb \} \phi_n^{\alpha*} \label{10} 
\end{equation} Then the integral (\ref{9}) acquires the form 
 \begin{equation} I= \int \exp \{  \sum_{\alpha} \{b \sum_{n=0}^{N-1} 
[b^{-2}( \phi^{* \alpha}_{n+1} \phi^{\alpha}_{n}+h.c.-2 \phi^{* \alpha}_n 
\phi^{ \alpha}_n) - \label{11} \end{equation} $$ -L^{1/2}( \phi^{\alpha*}_n
 \exp \{-(m-iD^{\alpha})bn \}(m-iD^{\alpha}) \chi^{\alpha}+h.c.]
 $$
 $$+ \frac{L}{2m} \chi^{\alpha*} \chi^{\alpha} \} \} d 
\phi^{\alpha*}_nd \phi^{\alpha}_nd \chi^{\alpha*}d \chi^{\alpha} $$ Now 
the quadratic form in the exponent does not depend on $D^{\alpha}$ and to 
calculate the integral it is sufficient to find a stationary point of the 
exponent. For small $b$ the sum over $n$ can be replaced by the integral 
 and the stationary equations by the following differential ones:  
\begin{equation} \ddot \phi^{* \alpha} -L^{1/2} 
  \chi^{*\alpha}(m+iD^{\alpha}) \exp \{-(iD^{\alpha}+m)\}=0\label{12} 
\end{equation} $$ \ddot \phi^{ \alpha}-L^{1/2} \chi^{ 
  \alpha}(m-iD^{\alpha}) \exp \{(iD^{\alpha}-m) \}=0 $$ $$ 
 \phi^{\alpha}(L)= \phi^{\alpha}(0)=0, \quad \phi^{\alpha*}(L)= 
\phi^{\alpha*}(0)=0 $$ The solution of these eq.s is \begin{equation} 
 \phi^{* \alpha}_{st}= L^{1/2} \frac{\chi^{* \alpha} \exp 
\{-(m+iD^{\alpha})t \}}{(m+iD^{\alpha})}+ \label{13} \end{equation} $$ 
+L^{-1/2} \frac{ \chi^{* \alpha}(t-L)}{m+iD^{\alpha}}-L^{-1/2} \frac{t 
\exp \{-(m+iD^{\alpha})L \}}{(m+iD^{\alpha})} $$
\begin{equation}
 \phi^{\alpha}_{st}=L^{1/2} \frac{ \chi^{ \alpha} 
\exp \{-(m-iD^{\alpha})t \}}{(m-iD^{\alpha})}+
\label{14}
\end{equation}
$$
+L^{-1/2} \frac{ \chi^{\alpha}(t-L)}{(m-iD^{\alpha})}-L^{-1/2} \frac{ 
\exp \{-(m-iD^{\alpha})L \} \chi^{\alpha}}{(m-iD^{\alpha})} $$  
  Substituting this solution into eq(\ref{11}) and integrating over $t$ 
we get  
\begin{equation}
 I=  \int \exp \{- 
\sum_{\alpha}  \frac{ \chi^{\alpha*} 
\chi^{\alpha}}{m^2+(D^{\alpha})^2} \} d \chi^{* \alpha} d \chi^{\alpha} 
+O(e^{-mL})
\label{15}
\end{equation}
 Integrating over $ \chi$ and omitting exponentially small 
corrections one gets 
\begin{equation} 
I= 
\prod_{\alpha}(m^2+(D^{\alpha})^2)= \det(-(\hat{D})^2+m^2)= 
\det( \hat{D}+m)^2  \label{16}
\end{equation}

As was discussed above the integral in the r.h.s. of eq.(\ref{6}) is a 
linearized version of the integral $I$, the differences being 
$O(b^2/a^2)$.
At the same time it is easy to show that replacing the sum in
the eq.( \ref{11}) by the integral over $t$ also produces  corrections of
order $O(b^2a^{-2})$. 
Therefore we proved that the square of quark determinant indeed can be 
presented as the path integral of the exponent of bosonic effective action 
(\ref{4}). For a finite lattice the corrections are of the order 
\begin{equation}
O(b^2/a^2)+O( e^{-mL})
\label{17}
\end{equation}

In distinction of the previous version \cite{AS2} the effective action 
described here does not include constraints, whch makes easier it's 
Monte-Carlo simulations. At present such simulations for simplified lower 
dimensional model are in progress.

{\bf Acknowledgements.} \\
This researsh was supported in part by Russian Basic Research
Fund under grant 96-01-00551.$$ ~ $$  \end{document}